# High resolution depth profiling using near-total-reflection hard X-ray photoelectron spectroscopy

Julien E. Rault [1,a)], Cheng-Tai Kuo [2], Henrique P. Martins[3,4], Giuseppina Conti [3,4], and Slavomír Nemšák[4]

[1]Synchrotron SOLEIL, L'Orme des Merisiers, Saint-Aubin-BP48, 91192 Gif-sur-Yvette, France
[2]Stanford Synchrotron Radiation Lightsource, SLAC National Accelerator Laboratory, Menlo Park, California 94025, USA
[3]Department of Physics, University of California Davis, Davis, California 95616, USA
[4]Advanced Light Source, Lawrence Berkeley National Laboratory, Berkeley, California 94720, USA

a) Electronic mail: julien.rault@synchrotron-soleil.fr

By adjusting the incidence angle of incoming X-ray near the critical angle of X-ray total reflection, the photoelectron intensity is strongly modulated due to the variation of X-ray penetration depth. Photoelectron spectroscopy (PES) combining with near total reflection (NTR) exhibit tunable surface sensitivity, providing depth-resolved information. In this review article, we first describe the experimental setup and specific data analysis process. We then review three different examples which show the broad application of this method The emphasis is on its applications to correlated oxide heterostructures, especially quantitative depth analyses of compositions and electronic states. In a last part, we discussed the limitations of this technique, mostly in terms of range of samples which can be studied.

# I. INTRODUCTION

Since its development in 1957,[1] photoelectron spectroscopy is a crucial technique to characterize materials of all kinds. All started with core-levels spectra of solid-state materials, a copper sample,[1] but it has quickly moved on to gas[2] and more recently to liquid systems.[3,4] This paper focuses on solid state materials and specifically buried interfaces. By measuring the kinetic energy of photoemitted electrons ($E_{KE}$), photoelectron spectroscopy intends to trace back to the initial binding energy of electrons ($E_{BE}$) in matter via a simple law of energy conservation $E_{BE} = h\nu - E_{KE}$, with $h\nu$ being the energy of the incident photons. All the subtlety of PES lies in how one interprets the binding energy. In the most general way, BE is the difference of energy between a system with N electrons in the ground state and an excited system with (N-1) electrons. In the simplest way, we use the frozen-orbital approximation and assume BE is simply the Hartree-Fock orbital energy of the orbital. Reference [5] describes most of the different approximations researchers make in PES experiments to link the measured kinetic energy to relevant physical information on the sample.

Over the years, PES has expended its capabilities with instruments capable of measuring all kinds of useful parameters. By detecting electron emission angles, one can use momentum and energy conservation to transform kinetic energy, emission angle into a spectral function in wavevector and binding energy. This measure is closely related to the ground state band structure of matter. This technique is commonly known as angle-resolved photoelectron spectroscopy (ARPES) and has allowed tremendous breakthrough in the understanding of crystalline solid state materials and its related properties such as metal-insulating transition, superconducting states, highly-correlated phenomena and so on.[5,6,7,8,9] The discrimination of the spin of outgoing electrons can be obtained in ARPES instruments so that scientists have a complete picture of spin-resolved band structure of materials, expending the knowledge of strong spin-orbit coupled materials, magnetic devices.[10,11,12] In parallel, developments allowed the measurements of the spatial location of emitted electrons, adding few 10 nm lateral resolution.[13,14,15,16] Finally, very recent improvements in vacuum techniques and instrumentation allowed PES to overcome one of its main limitations: the need of ultrahigh vacuum. Even though the first pioneering experiments of Siegbahn and his group on gases[17] and liquids[18] took place already over 50 years ago, near-ambient-pressure PES has only recently become commonly available in laboratories all over the world[19] and is the next big step forward to better understand materials in realistic conditions such as lithium batteries and catalytic materials.[20]

Professor Charles S. Fadley, friends called him Chuck, was a key player in all the aforementioned improvements of photoelectron.[21] He was also very involved in the use of PES on synchrotron source facilities, and he was one of the leading figures in taking advantage of the outstanding brightness and full energy range of synchrotron radiation to conduct photoelectron in the hard X-ray regime.[22] Working with hard X-rays carries photoelectron over one of its main limitations: a very low probing depth. Along with others,[23,24,25,26,27,28] they practically established the field of hard X-ray photoelectron emission spectroscopy (HAXPES) and the "bulk" PES, by increasing the photon energy from soft X-rays (typical lab based XPS 1.486 keV) to several keV in HAXPES. Higher energy photoelectrons, with a greatly higher inelastic mean free path (IMFP) ranging up to 10 nm,[29] then allow for probing much deeper. It is well known that the photoionization cross-sections decrease dramatically when excitation energy is increased from soft to hard X-ray regime. The rate of the drop in the cross-section values is, however, not the same for all electronic levels. It is therefore, on top of using Cooper minima, or polarization dependence, another way to enhance or suppress the signal from a specific core-level or a valence band state. Broad overview of HAXPES, including details on photoionization cross-section can be found in a recent review[30].

One of Chuck's main efforts withing HAXPES X-ray was to access depth-resolved information, and he focused a lot of his work to angle-resolved PES.[31] In a classic PES experiment, photons impinge the surface at an angle of ~45 deg, hence penetrating over several µm in the sample (see Fig. 1). To get depth-resolved information in this geometry, one has to take advantage of the escape depth of emitted electrons, and its variation as a function of the electron emission angle[32] or as a function of its kinetic energy.[33] Analyzing such variable-angle or -energy data clearly can provide some indication of the relative depths of different species, but the analysis involves an ill-determined convolution integral that may need additional compositional constraints to be solved.

Following the early experiments of Batterman in fluorescence yield,[34] C. Fadley and others (W. Drube, J. Zegenhagen, M. Bedzyk) combined the X-ray standing waves with photoelectron spectroscopy. By doing that, they developed another method to get depth resolution with photoelectron: instead of taking advantage of the variable IMFP of electrons, they used X-ray optical interference effects to modulate the photon field inside the sample to get electrons excited from varying depth. Such X-ray interferences effects at interfaces are extensively described in the following reference.[35]

The first approach to achieve such modulation is to use reflection off the single-crystal atomic planes to create a standing wave inside (and above) the material.[36] By either changing the incidence angle or photon energy across the Bragg value, the nodes and antinodes of the wave will shift in the direction perpendicular to diffraction planes by half of the standing-wave period, making possible a very precise control of photoemitted electrons provenance. Probing buried interfaces by Bragg-reflection XPS using multilayer mirrors has been pioneered by the Fadley group[37,38] and it is a main subject of another paper in this Special Issue.[39] The main drawback of this method is the need of periodic multilayer samples, which increase the complexity of the sample synthesis, or even cannot be used for some asymmetric systems as we'll discuss below.

The main interest of this paper is then to describe a second approach to modulate X-ray wavefield intensity: the use of photon beams close to critical angle in a near-total reflection mode, ultimately to get depth-resolved information on its electronic and chemical properties. Near-total-reflection X-ray photoelectron spectroscopy (NTR-PES) was first proposed by Henke in 1972, who demonstrated that the photoelectron intensity at the surface is maximized (quadrupled in the antinodes of the formed standing wave) at the critical incidence angle.[40] In the very early stages of this technique, Fadley's group contributed several important experimental and theoretical findings. His group reported enhancement of the surface PES intensity at NTR condition on various materials and developed its theoretical model for data analyses.[41,42,43] In 90s, J. Kawai et al. continued to develop this method and show the additional advantages of NTR-PES, such as reduction of inelastic backgrounds, tunable surface sensitivity, and analysis of Kiessig fringes.[44,45,46] The detailed history of the development of NTR-PES technique can be found from prior review papers.[47,48] In early 2000s, NTR-PES was recognized as a non-destructive, surface analytical method of determining the surface contaminations or overlayers, which is especially valuable to the semiconductor industry.[49,50] In principle, through a slight change of the incidence angle around the TR critical angle, the probing depth of HAXPES can be monotonically varied. In the other words, the photoelectron intensity around the TR critical angle (typically 0~3°) is strongly affected by the rapid change of X-ray penetration depth. If all the relevant optical constants of the solids are known, the NTR-HAXPES intensity can be analyzed, and the buried depth information can be extracted. In addition to the strong increase of photoemission yield at such low angles, this technique can serve as useful solution when classic depth-resolved techniques are not feasible. Photon energy scans cannot be conducted in laboratory X-ray tube-excited HAXPES instruments, and varying emission angle to a very shallow emission can be problematic as well (for example in ambient pressure versions of hemispherical analyzers with entrance cones). Even at synchrotron facilities, one might want to use one specific energy, to take advantage of electronic resonances or to prevent changes in energy shift or beam movements. Lastly, when probing buried interfaces or other depth gradients, the main advantage of HAXPES is not only its increased information depth, but the tunable depth selectivity which can be successfully achieved, among other ways, by using X-ray optical effects near total reflection. We will demonstrate these benefits in the following sections. It is important to state that the surface and interfaces of the sample should be planar (as opposite to e.g. porous) to fully benefit from the information depth selectivity of NTR-HAXPES. Otherwise the incident angle of photons is ill-defined, leaving the only feasible option of depth selectivity via varying excitation energy.

Note that in addition to the term of NTR-PES, in some studies it has been used as grazing incidence XPS (GIXPS) or total reflection XPS (TRXPS), but in this article we will use NTR-PES. NTR-PES has gradually been applied to various kind of solid state materials, such as semiconductors,[51,52] nano materials,[53] correlated oxides,[54,55,56] and also extreme UV light photoresists.[57]

In this paper, we will explore a few recent NTR-HAXPES works to help describe how this powerful method can be used in practice on any sample. In the section II, we will introduce some of the practical aspects of performing NTR-HAXPES experiments, including data collection and analysis. In the sections III-V, examples of using NTR-HAXPES to probe the surface and bulk electronic structure of correlated oxides,[55] the functional interface between two perovskite oxides ($BiFeO_3/Ca_{1-x}Ce_xMnO_3$ bilayer[54] and $BiFeO_3/La_{0.7}Sr_{0.3}MnO_3$ superlattice[56]) are discussed in detail. In the last section, we will discuss specific resonant effects in NTR photoelectron, and how they can help with application of this technique to a broad variety of systems.

## II. NTR-PES methodology, data acquisition and analysis

Conventionally, HAXPES experiments use a normal electron emission with a grazing beam incidence (a few degrees above the critical angle). In this geometry, the penetration depth of photons is much larger than the escape depth of photoelectrons and the probing depth of the HAXPES is governed by the escape depth of photoelectrons.

On the other hand, if the incidence angle of X-rays is set below the critical angle of total reflection, the X-ray penetration depth decreases significantly, and it may become shorter than the escape depth of photoelectrons. The effective X-ray attenuation length ($\Lambda_{eff}$) of a solid depends on its index of refraction, and the X-ray incidence angle. The attenuation lengths of X-rays can be readily obtained using CXRO database.[58] In this case, the probing depth of NTR-HAXPES is governed by the X-ray penetration depth, leading to enhanced surface sensitivity. In terms of length scales, the probing depth of NTR-HAXPES can be as low as ca. few nanometers while that of conventional HAXPES can be as high as several tens of nanometers.

Since critical angles of solids in multi keV regime vary between ~ 0.1 to ~2°, precise scanning of the incidence angle is employed to vary the information depth in NTR-HAXPES experiments. Sample stages are usually mounted on rotatable goniometers, which can provide reproducible polar rotation of the sample with accuracy on the order of 0.01°. Scans usually cover ranges from 0° incidence (beam parallel to the sample surface) to high enough incidence angles (past the critical angle) to see the oscillations from the formed X-ray standing wave. Photoelectron core-level intensities of interest are recorded for each incidence angle and the resulting intensity profiles as a function of. incidence angles are defined as rocking curves (RCs). It is important to mention that the photoelectron intensities versus incidence angles are also called photoelectron yield in the SW community.

An important aspect of data analysis involves X-ray optical and photoelectron emission simulations. Yang X-ray Optics (YXRO) program, developed by Fadley group in early 2000s, has been made available to the user community.[59,60] It is a powerful tool for the quantitative depth analyses, that can be utilized not only to analyze experimental data after the measurements, but also to simulate the theoretical NTR-PES intensities for various sample structures before the actual experiments are carried out to help with the experiment design. Another, more recent addition to the analysis tools of the NTR-HAXPES is called SWOptimizer.[61] It is an optimization algorithm that uses YXRO backend for simulating theoretical rocking curves based on a structural model of the sample. SWOptimizer then iteratively changes the parameters of the model (such as thicknesses of individual layers, or roughnesses of interfaces) to gain best fit between the experimental and simulated data. Program uses a black-box optimizer and a surrogate model to increase the speed of the optimization and increase the robustness against finding false local minima.

## III. (N)TR-HAXPES to probe electronic structure of correlated oxides

HAXPES made a huge impact in studying bulk electronic structure of solid-state materials. Its main feature is the large probing depth in comparison with the soft X-ray PES. The probing depth of normal HAXPES (labeled by Mizutani et al.[55] as NR-HAXPES, see Fig. 2(a)) is governed by the escape depth of photoelectrons. NR-HAXPES generally uses a grazing incidence geometry (e.g. incidence angle of ~3°). For HAXPES in NTR regime (Fig. 2(a), labeled as TR-HAXPES), the incidence angle is set at the TR critical angle, which is ~0.3° for LSMO for ~8 keV photons used in this experiment. For the case of LSMO, the probing depth for TR-HAXPES is around several nanometers, which is much shorter than that of NR-HAXPES (several tens of nanometers)[55]. Utilizing the dramatically shorter probing depth of TR-HAXPES, Mizutani et al. demonstrated that they were able to observe both surface and bulk electronic states of solids.

Figure 2(b) shows Mn 2p core-level HAXPES spectra ($hv$ = 7.94 keV) of a $La_{0.7}Sr_{0.3}MnO_3$ (LSMO) thin film. The NR-HAXPES spectra agree well with prior results[62,63] with a clear shoulder (labeled as β) at ~638 eV. One can observe that feature β is suppressed in the surface-sensitive TR-HAXPES spectra, and this result shows the same qualitative behavior with respect to the previous soft X-ray PES results.[64,65] The feature β is known to originate from well-screened states that were explained by the configuration interaction model for a series of transition metal compounds,[58,63,62] and it is directly connected to the bulk electronic structures.

This finding directly demonstrates that the probing depth of HAXPES can be shortened by one order to be several nanometers by making use of a TR geometry. One can probe either surface- or bulk-sensitive electronic states of solids at a HAXPES endstation, depending on the interests of studies, which is a great advantage. To study the electronic structure of a solid, it is often necessary to compare the difference between the HAXPES and SX-PES spectra for distinguishing the bulk and surface electronic states. However, most photoelectron endstations don't cover wide range of photon energies to provide both HAXPES and SX-PES experiments at the same beamline. Because of this, the samples need to be measured separately in at least two different beamlines with usually different experimental conditions (e.g. energy resolution, measuring geometry, and sample conditions). The difference in experimental conditions of HAXPES and SX-PES beamlines could make direct comparison of PES spectra difficult. Simultaneous measurement of TR- and NR-HAXPES at the same beamline can resolve this issue and directly discriminate the surface electronic states from the bulk ones. Although TR-HAXPES is useful in discriminating surface and bulk features, it is important to point out its drawback that the typical energy resolution of several hundred meV for HAXPES may limit its potential in studying these enhanced surface features.

Another aspect is the difference between TR-HAXPES and take-off angle (TOA)-dependent HAXPES. TOA-dependent HAXPES is widely used for discriminating the surface and bulk states, but its change of probing depth is less sensitive compared to TR-HAXPES as they are governed by different natures. The minimum escape length of the photoelectrons in the TOA-dependent HAXPES is ~10 nm or more.[66] In contrast, the minimum probing length can reach a few nm for TR-HAXPES.[52] In summary, control of total reflection with HAXPES can tailor the probing depth of HAXPES, allowing simultaneous measurements of surface and bulk electronic properties of solids with exact same experimental conditions. This capability is useful for the future studies to eliminate controversies about surface and bulk states.

In the next section, we will introduce the methodology of quantitative depth analyses of depth-resolved NTR-HAXPES results.

# IV. Depth-resolved NTR-HAXPES: applications to a bilayer $BiFeO_3$/$Ca_{1-x}Ce_xMnO_3$ bilayer heterosturcture

This case study explores how ferroelectric polarization can trigger a metal-insulator transition in an adjacent material. A ferroelectric is an insulating system with two or more discrete stable or metastable states of different nonzero electric polarization in zero applied electric field, referred to as spontaneous polarization.[67] When ferroelectric polarization arises in the out-of-plane direction of a finite slab, it will induce a surface charge density at the surface. These charges generate an electric field, named depolarizing field, which acts against the ferroelectric polarization.[68,69] When put in contact with another material, the screening of these charges can have significant consequences on both the ferroelectric layer and the adjacent layer which will adjust its own electronic properties to maximize the screening.[70] This effect is particularly interesting when the adjacent layer is a material on the verge of physical transition which can be triggered by slight changes in doping.[71] Such ferroelectric control of functional properties is of particular importance in the case of Mott insulators, opening the possibility of the realization of smaller or less energy consumption switches or memory devices. A strong electro-resistance effect induced by the ferroelectric polarization switching has been demonstrated in heterostructures built with a ferroelectric $BiFeO_3$ (BFO) and such a Mott insulator based on the $Ca_{1-x}Ce_xMnO_3$ (CCMO) electron doped manganite.[72,73] To better understand this effect, charge distribution measurement on a unit cell scale at the CCMO/BFO interface is required to understand, control and possibly enhance the magnitude of the switching effects. The BFO/CCMO heterostructure sample was grown by pulsed laser deposition on a (001) $YAlO_3$ substrate, as sketched in FIG. 3(a). This study was initially achieved by the combination of scanning transmission electron microscopy (STEM) plus electron energy loss spectroscopy (EELS),[74,75,76,77] and NTR-X-ray HAXPES. Only the latter will be described here.

Since we want to tailor the field, the first step is to compute the electric field in the NTR to see whether we can extract any useful information. FIG. 3(a) shows the simulated X-ray intensity (hv = 2800 eV) in the BFO/CCMO. The incidence angle is scanned over the NTR range: from total reflection at zero incidence angle to 3°. A thin layer of carbon and oxygen containing species has been added on top of the bilayer to simulate the surface contaminants[55,83].

The calculation reveals that, for increasing incidence angle θ, the X-rays penetrate deeper into the bilayer. It goes from ca. 5 Å at 0.69° to 250 Å at 1.0°, with clear oscillations in photon field magnitude as a function the incidence angle. A schematic representation of the bilayer BFO/CCMO heterostructure is given in the FIG. 3(b). In FIG. 3(c) and 3(d) are shown the Ca 2p core-level spectra at incidence angles θ = 0.69° (more interface sensitive) and 1.5° (more bulk sensitive), respectively. Each member of the spin-orbit-split doublet shows two components with a binding energy (BE) difference Δ BE=0.70 eV. This indicates that the Ca atoms in the CCMO layer are in two distinct chemical, structural or electronic potential environments. The spectral weight of the high-binding-energy (HBE) component decreases relative to the low-binding-energy (LBE) one when going from θ = 0.69° to 1.5°. In FIG. 3(a) one can see that at θ = 0.5° the X-rays field only reaches the top of the CCMO layer while at θ = 1.5° the intensity is significant deeper into the sample. Then, we can safely allocate the HBE component to an interface region and LBE component to the bulk CCMO. More quantitative insight on the depth distribution of all the chemical components present in the structure, can be inferred by the analysis of the full set of NTR curves. Indeed, the X-ray intensity shows strong oscillations over incidence angles of about 0.9° to 2° as a function of the incidence angle. This beating pattern is related to the creation and vertical motion of a standing wave and can be used to selectively probe different depths. Scanning the incidence angle θ while measuring core-level photoelectron signal from each element in the sample over this NTR range, we generate intensity profiles produced by the varying X-ray intensity depth profile, with each spectrum carrying information on parameters such as spectral weight and binding energy position.

The experimental photoelectron NTR curves extracted from core-levels representative of each layer in the sample are shown in FIG. 3(e) (symbols). As expected from the electric-field calculation in FIG. 3(a), the RC curves clearly show significant intensity oscillations as a function of the photon incidence angle (e.g., ≈ 20% above the total-reflection falloff and located at 0.80° for the Bi 4f core-level). The oscillations are also phase-shifted when going from one layer to another (0.20° when going from BFO to bulk CCMO), which clearly demonstrates the incidence-angle dependence of the probing depth provided by the standing-wave phase information in the field profile.

To obtain a quantitative depth profile of the structure we compare the experimental NTR curves to calculations done with the YXRO program for photoelectron. The plain lines of FIG. 3(e) show the simulated photoelectron yield for a theoretical heterostructure. The best fit to the experimental curves is thus obtained for interface CCMO layer with thickness 10 (i.e. 2.5 u.c.). In this study, the only free parameters were the interface thickness. Other thicknesses have been fixed to the values expected from the growth. Thus, the structure that best describes the heterostructure consists in a top $CO_x$ contamination layer (10 Å), a BFO layer (with thickness of 42 Å) and a CCMO layer with an interface part (10 Å) and a bulk part (210 Å). These optimal thicknesses determined from the calculation are summarized in FIG. 3(b). This result is in good agreement with the results obtained by complementary STEM-EELS,[54] which show a highest electron concentration over the first two-unit cells close to the interface with the ferroelectric $BiFeO_3$, as expected for a downward polarization.

In most compounds, electron-doping induces a shift towards lower BEs. For instance, in $SrTiO_3$, electron doping induced by oxygen vacancies leads to the appearance of a $Ti^{3+}$ peak at lower BE than the $Ti^{4+}$ peak.[78] Nevertheless, in $ABO_3$ compounds, one would expect a different behavior between the covalent $BO_2$ and the ionic AO layers. Vanacore et al. showed this is indeed the case in $SrTiO_3$ single crystals.[79] The surface-related HBE peak of Sr 3d is due to a lower coordination, or equivalently a higher electron occupation, at surface cation sites. The same behavior has been observed in $BaTiO_3$.[80] Van der Heide showed that Ca, which is in the same periodic table column as Sr and Ba, behaves similarly i.e. that a higher electrons occupation induces higher-binding-energy for Ca core-levels.[81] Finally, Taguchi and Shimada showed that oxygen vacancies induce a HBE peak in $CaMnO_3$.[82] All of this suggests that the HBE component we see in FIG. 3(c) is indeed due to a local electron doping of the $Ca_{1-x}Ce_xMnO_3$ layer. The NTR-HAXPES results are thus in agreement with the charge accumulation at the $Ca_{1-x}Ce_xMnO_3$ interface obtained by the STEM-EELS

This first example gives a good illustration of the capabilities of the NTR technique alone in non-multilayer systems. NTR can also be used in combination with SW on multilayers. It will be the aim of the next example. In this example, we tackled another key aspects of oxide heterostructures: interface quality. Indeed, a key parameter influencing the emergent properties in heterostructures is the quality of interfaces, where varying interdiffusion lengths can give rise to different chemistry and distinctive electronic states.

## V. Depth-resolved NTR-HAXPES: applications to a $BiFeO_3$/$La_{0.7}Sr_{0.3}MnO_3$ superlattice

Thanks to its complex electronic and magnetic properties, there have been numerous in-depth studies of the manganite perovskite $La_{0.7}Sr_{0.3}MnO_3$ (LSMO).[83,84,85] As in many other perovskites, LSMO magnetic and electronic transport properties depend on the film thickness and also on the used growth techniques.[86]

Growing LSMO on top of a ferroelectric compounds, such as BiFeO$_3$ (BFO) will also have a significant effect on its physical properties. Epitaxial growth of magnetic layers on ferroelectric materials make "hybrid multiferroics" with crucial potential applications in the field of low-energy electronics, since it would allow the control of magnetization by an electric field.[87] The BFO/LSMO heterostructure is one of the candidates to study coupling in the hybrids multiferroic material. In such systems, a key question is whether new electronic or magnetic states emerge at the interfaces between BFO and LSMO. Interdiffusion at the interfaces plays a major role in magnetic heterostructures, where rough or sharp interfaces give rise to different magnetic properties.[88,89] Therefore, it is essential to have depth-resolved information at such interfaces. BFO and LSMO heterostructures have been thoroughly studied and their electronic structure has been studied by conventional spectroscopy techniques and by first principles theoretical approaches.[87,88,90,91]

H.P. Martins et al.[56] studied a BFO/LSMO superlattice fabricated on a TiO$_2$-terminated (001) single-crystal SrTiO$_3$ substrate by pulsed laser deposition (FIG. 4(a)). By using Bragg reflection SW-HAXPES, two interfaces in the superlattice (BFO on top of LSMO and LSMO on top of BFO) were found to have dramatically different properties. The interfaces where BFO is on top are sharp and abrupt. In contrast, the interfaces with LSMO on top exhibited 1.2 u.c. (4.8 Å) interdiffusion length.

However, interesting information on the surface properties can be obtained using the NTR region. FIG. 4(b) shows the calculated electric field strength as a function of sample depth and X-ray incidence angle. FIG. 4(c) shows the photoelectron yield of different core-level electrons as a function of incidence angle. The order of the onsets for each core-level follows the depth order of the probed layers. The onsets for the Bi 5d and O 1s signals rise first, showing the sample surface is BiO-terminated. At slightly higher incidence angles, signals from Fe 2p, Sr 2p take off, and finally, La 3p and Mn 2p intensities start to rise. Thanks to distinctive chemical states in Bi 5d, O1s and Sr 2p, the photoelectron onset curves were separated into low- and high-binding energy components of these core-levels. FIG. 4(d) and FIG. 4(e) show the Bi 5d and O 1s curves with clear different onsets for both components, indicating that the high-binding energy (higher oxidation state of Bi) component comes from the surface layer. FIG. 4(f) shows the high-binding energy component of the Sr 2p core-level comes from deeper than the main low-binding energy component.

Thanks to NTR, we get a bit more information: on one hand, the abrupt BFO/LSMO interface has a higher Sr concentration. This region is associated with the HBE component in Sr 2p core-level, meaning it's in a different chemical environment and contained within 1 unit cell from the interface. This region might be associated with the magneto-electric coupling. On the other hand, the top BFO layer contains exclusively Bi atoms with a HBE component in Bi 5d core-level spectra. This is probably due to air exposure of the sample, where surface Bi in BFO can be further oxidized. The buried BFO layers contain the stochiometric Bi atoms with the LBE Bi 5d signature.

In summary, this example demonstrates the complementarity of the NTR-HAXPES Bragg reflection SW-HAXPES when investigating a superlattice sample. SW-HAXPES (thanks to its higher information depth) probes deeper through several buried interfaces of the superlattice, while NTR-HAXPES focuses on the few topmost interfaces. Since samples are often exposed to air before the HAXPES, the surface layers of the superlattice are often vastly different from the rest of the sample. Combing NTR- and SW-HAXPES gives a more precise picture of the interfaces of the buried layers within the superlattice by discriminating the information coming from the contaminated surface.

## VI. Additional discussions on resonant effects

On this final section, we want to highlight a feature that one must have in mind before conducting NTR- or SW-PES experiments. The few examples described before show strong oscillations in NTR PES thanks to very different optical properties in the hard X-ray range between the adjacent layers. In the soft/tender X-ray range, BFO is optically significantly more dense than LSMO or CCMO. The high optical contrast between individual layers in this heterostructure then leads to enhanced reflectivity and strong interferences. Unfortunately, this technique is less efficient when the two adjacent layers have similar optical properties. For instance, FIG. 5(a) shows the theoretical PE yield of a model heterostructure made by 5 nm ferroelectric $BaTiO_3$ (BTO) layer grown on top 20 nm metallic $SrRuO_3$ (SRO). The interferences are weaker and the PE yield for the two relevant core-levels, namely Ba *3d* for the BTO overlayer and Sr *3d* for the SRO underlayer, show very weak standing wave oscillations difficult to use for quantitative analysis. This does not mean one should give up NTR- or SW- PES when adjacent layers have close optical indexes. Using resonant conditions for cleverly chosen photon energy, we can enhance the refraction and obtain significant oscillations even in materials which have close optical indexes "off-resonance". FIG. 5(a) shows PE yield at 2000 eV photon energy (i. e. close to the $L_2$ resonance for Sr) and at 2800 eV (off resonance), indicating a much stronger effects close to the resonance, revealing several oscillations (in resonant conditions).

In another example, FIG. 5(b) shows the experimental PE yield of Ir *4f* core-level in a SrIrO$_3$/LaFeO$_3$ multilayer as we scan through the La M$_5$ resonance (hv ~ 833 eV). We clearly see a dramatic enhancement of the Bragg feature and Kiessig oscillations in this SW-PES experiment when we are getting closer to the resonance, with a maximum effect just before the edge at hv = 831 eV. Another interesting effect is the standing wave phase reversal for hv = 833 eV, just on the falling-off slope of the M$_5$ absorption peak. This can be observed in the shape of the Bragg feature and between red and blue curves – position of minima and maxima is reversed. As the excitation energy gets even higher above the absorption edge at hv=840 eV, the Bragg and Kiessig oscillations become virtually invisible. Experimentally, Fadley group took advantage of these resonant effects in numerous Bragg-reflection standing wave studies.[92,93,94,95,96,97]

## VII. CONCLUSIONS

In conclusion, NTR is a powerful addition to HAXPES to allow depth-resolved measurements. Using three highlights, we showed how this technique can give precise, non-destructive information on the chemical profile of non-multilayered heterostructure on one hand, but also can improve the analysis of Bragg reflection SW--HAXPES studies of superlattices by providing complementary information. This technique can be used successfully on any structure where the adjacent layers have different optical densities, which allow for creating of strong standing wave in NTR regime, as showed in our last part. However, tuning the photon energy around resonance can change dramatically a material's optical index and make this technique usable on virtually any heterostructures. A careful theoretical study and analysis using YRXO is crucial beforehand any actual experiment. Another numerical tool, SWOptimizer, can then help with fitting the experimental and simulated data, and with automation of the search for a correct structural solution. Finally, as more and more beamlines are capable of using both soft and hard X-rays on the same experimental stage, adding soft X-ray or ultraviolet photoelectron spectroscopy with very high energy resolution can be of great use to detect subtle surface effects not seen with NTR-HAXPES.

# ACKNOWLEDGMENTS

The experiments on BFO/CCMO used a HAXPES end-station at beamline 9.3.1 of the Advanced Light Source, LBNL Berkeley (USA), a U.S. DOE Office of Science User Facility under contract no. DE-AC02-05CH11231. H.M. was supported for salary by the U.S. Department of Energy (DOE) under Contract No. DE-SC0014697. Use of the Stanford Synchrotron Radiation Lightsource, SLAC National Accelerator Laboratory, is supported by the U.S. Department of Energy, Office of Science, Office of Basic Energy Sciences under Contract No. DE-AC02-76SF00515. JER acknowledges the support of a public grant from the "Laboratoire d'Excellence Physics Atom Light Matter" (LabEx PALM) overseen by the ANR as part of the "Investissements d'Avenir" program (reference: ANR-10-LABX-0039). We acknowledge SOLEIL for provision of synchrotron radiation facilities. We thank P. Le Fèvre and D. Céolin for useful discussions.

## DATA AVAILABILITY

Data sharing is not applicable to this article as no new data were created or analyzed in this study.

## LIST OF FIGURES

FIG. 1. Typical XPS experiment with 45 deg photon incidence angle. Depth resolution is obtained by varying the electron take-off angle. One is more sensitive to the surface for low take-off angle (25 deg) (right panel) compared to normal emission (left panel). Inset shows the O 1s core-level of the sample with a surface-related component (high-binding energy peak) and a bulk-related component (low-binding energy peak).

FIG. 2. (a) Probing depth of normal HAXPES (NR-HAXPES) (left) and total reflection HAXPES (TR-HAXPES) (right). (b) Comparison of Mn 2p core-level spectra between NR-HAXPES and TR-HAXPES from a LSMO thin film. Experiments have been done at hv = 7940 eV Reprinted with permission from T. Mizutani et al., Phys. Rev. B 103, 205113. Copyright 2021 American Physical Society.

FIG. 3. (a) Calculation of the depth-resolved electric field strength |E2| as a function of depth and photon incidence angle for hv = 2800 eV. This demonstrates that changing the incidence angle permits selectively probing different depth. The oscillatory features represent standing-wave formation. (b) Schematic of the $BiFeO_3/Ca_{1-x}Ce_xMnO_3//YAlO_3$ heterostructure, with the thicknesses of the layers obtained after

fitting the experimental rocking curves by using the YRXO software[599] also indicated. Ca 2p core-level spectrum taken at incidence angles of (c) θ = 0.69° and (d) θ = 1.5°. The decrease of the HBE component when going from incidence angle 0.69° (interface) to 1.5° (bulk) suggests that this component belongs to the interface layer. (e) Experimental (blank circles) rocking curves for C 1s, Bi 4f, interface Ca 2p and bulk Ca 2p core-levels, overlaid with the calculated curves. Reprinted with permission from Marinova et al., Nano Letters 15, 2533. Copyright 2015 American Chemical Society.

FIG. 4. (a) The experimental geometry and sample structure of a BFO/LSMO superlattice. (b) Calculated electric field intensity as a function of sample depth and incidence angle near the TR region. (c) NTR-HAXPES intensities of different core-levels. NTR-HAXPES intensities for low- and high-binding energy components of (d) Bi 5d, (e) O 1s, and (f) Sr 2p core-level spectra. All experiments have been done at hv = 5953.4 eV. Reproduced with permission from H. P. Martins et al., arXiv:2012.07993 licensed under a CC BY 4.0 copyright.

FIG. 5. (a) Calculated PE yield for a model $BaTiO_3/SrRuO_3/SrTiO_3$ heterostructure for Ba 3d core-level (in black), and Sr 3d (in red). The calculations have been done at hv = 2000 eV, i. e. close to the L2 edge of Sr (plain lines) and at hv = 2800 eV, i. e. far from any resonance (dashed line). Inset shows the model heterostructure composition. (b) Experimental PE yield for Ir 4f core-level of a $SrIrO_3/LaFeO_3$ multilayer. Figures shows the evolution of SW oscillations as a function of hv, close to resonance (831.6 eV in blue, 832 eV in black, and 835 in red curves) and far from resonance (840 eV, green curve). Inset shows the experimental La $M_{4,5}$ absorption edge.

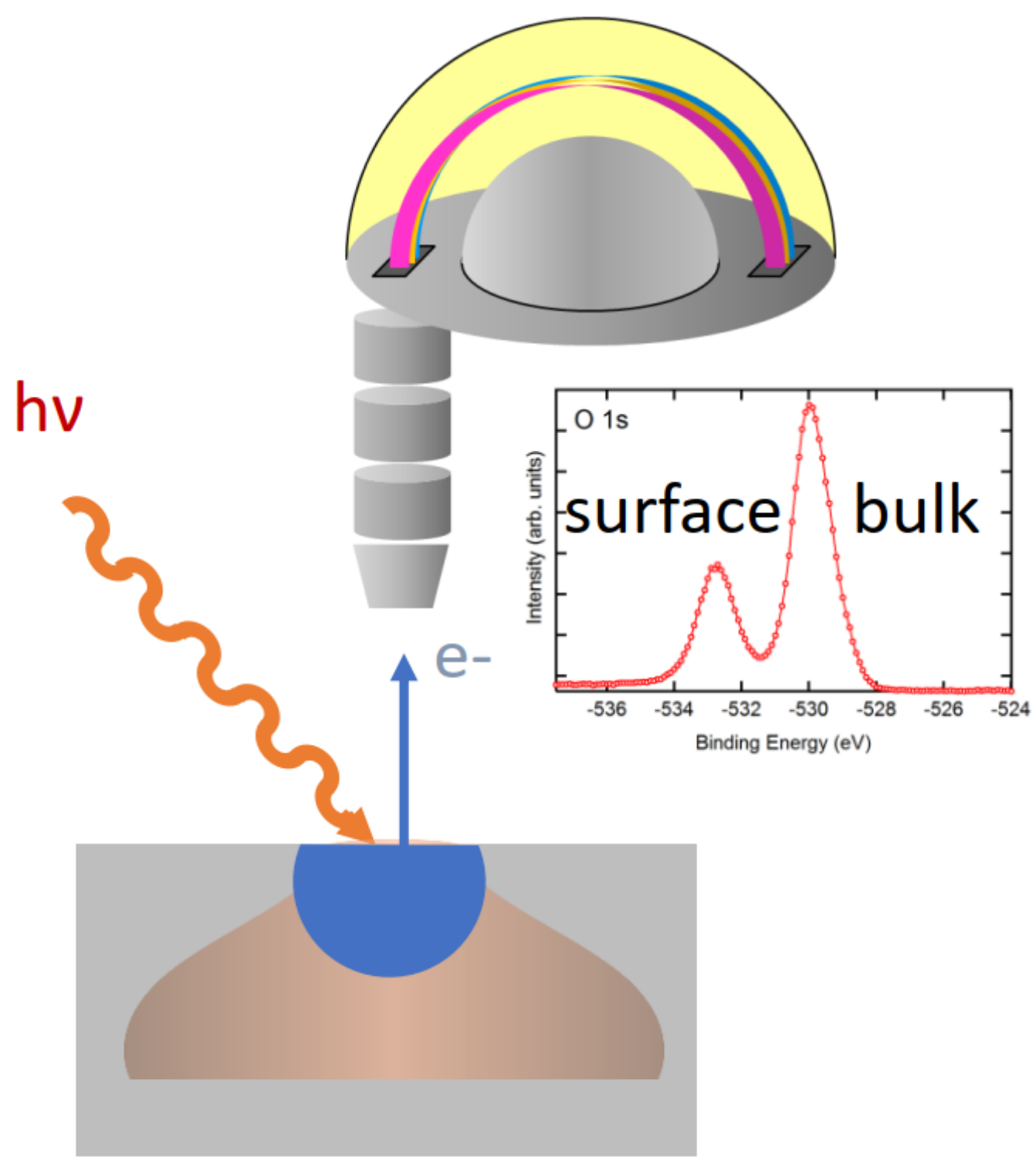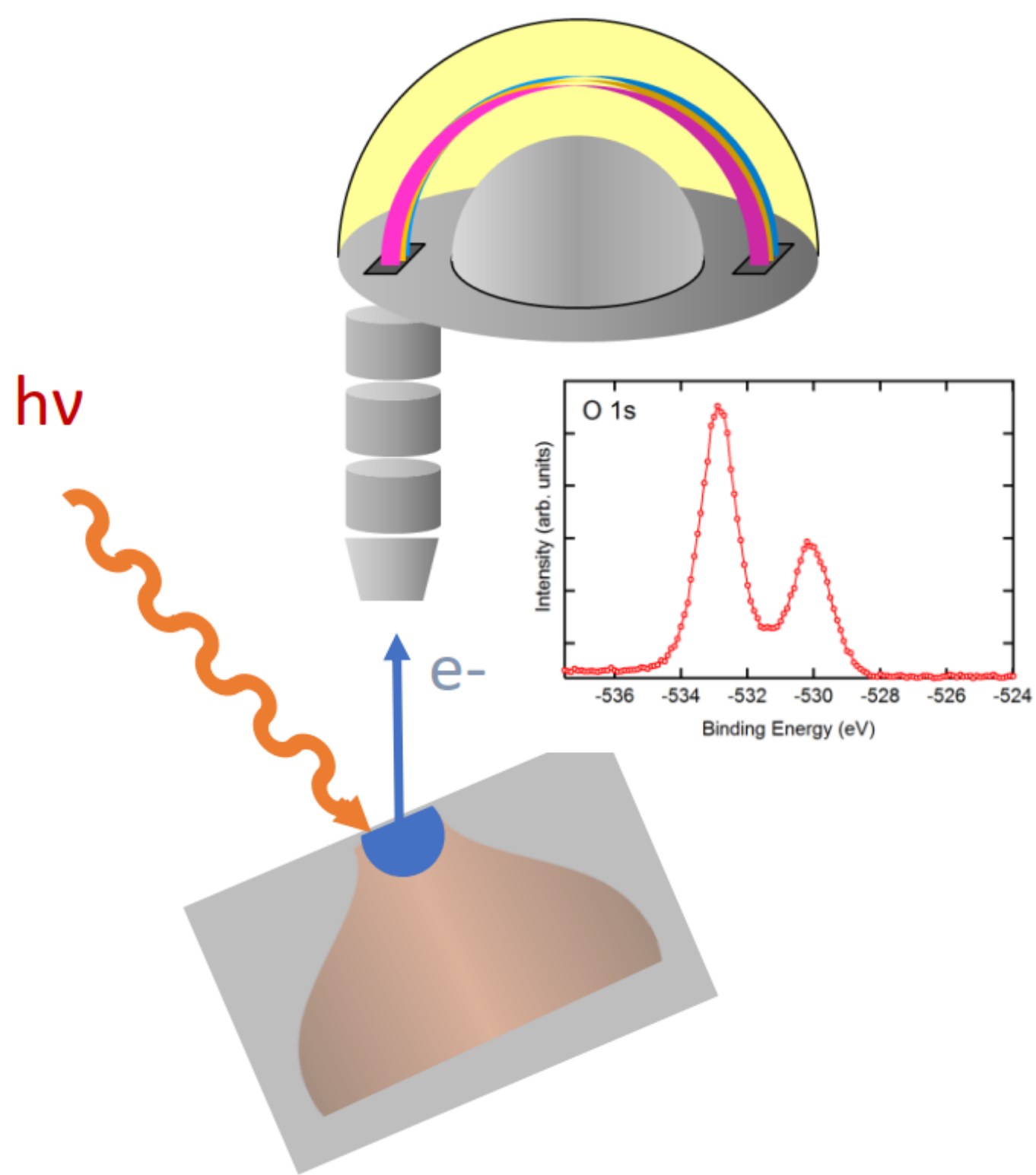

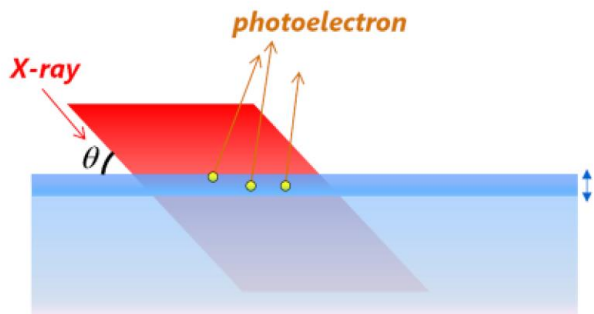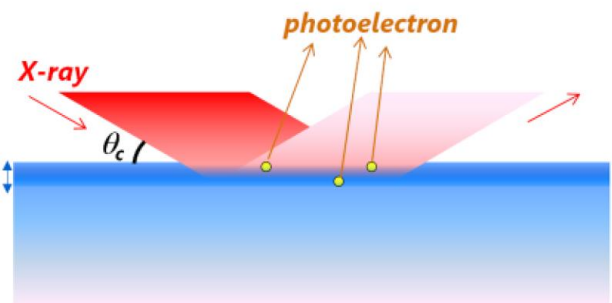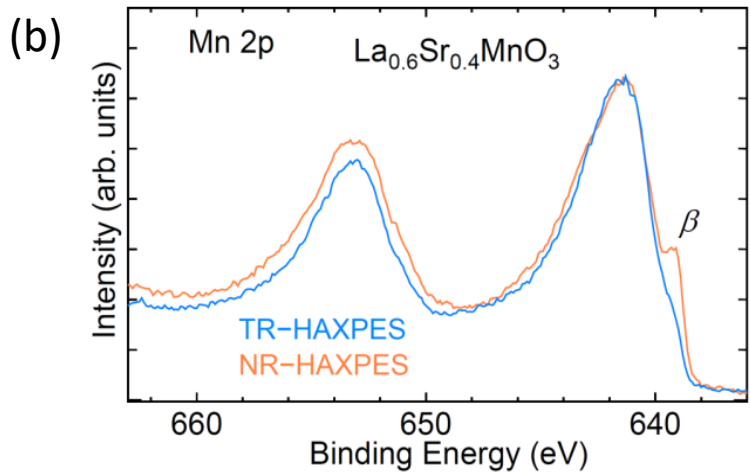

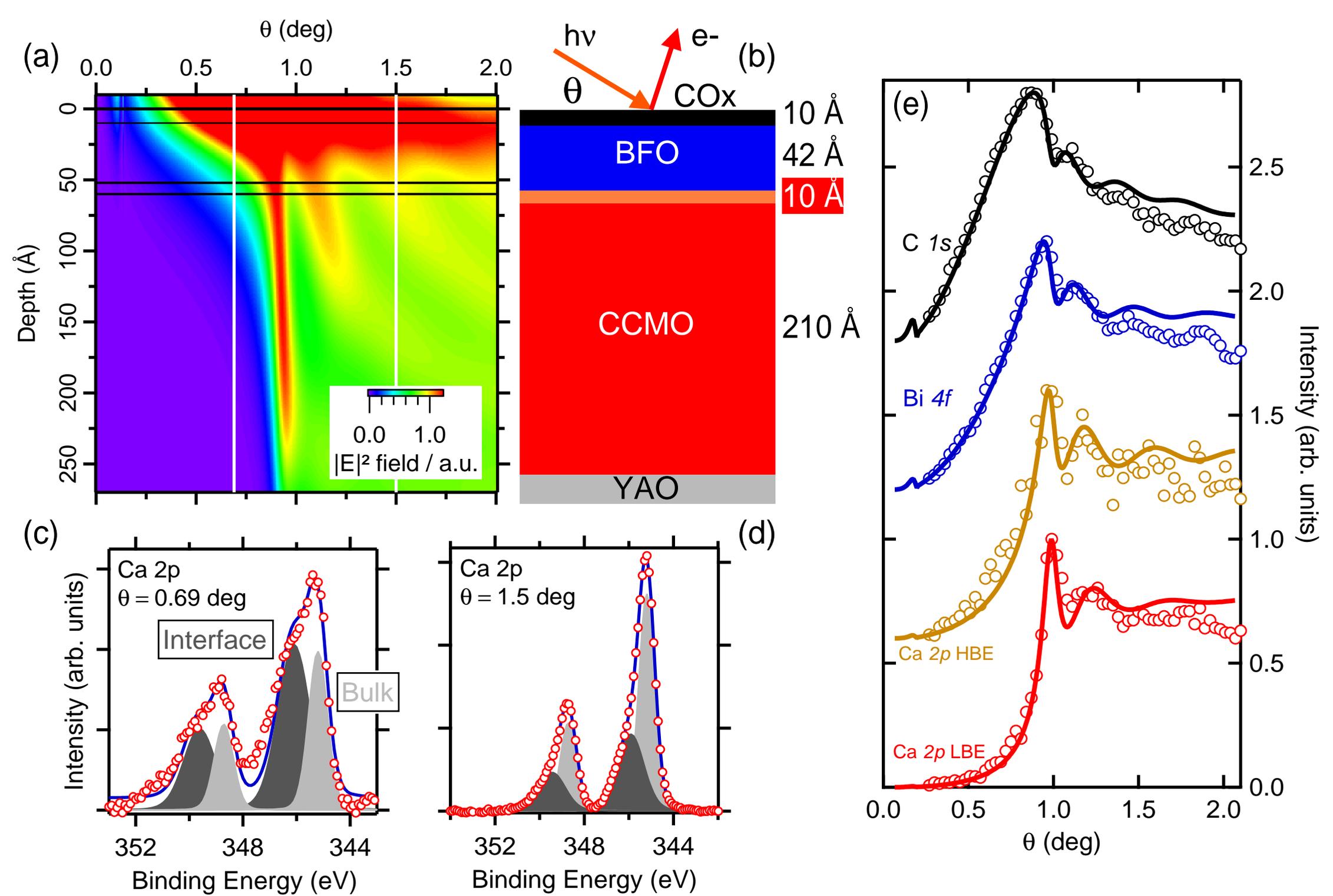

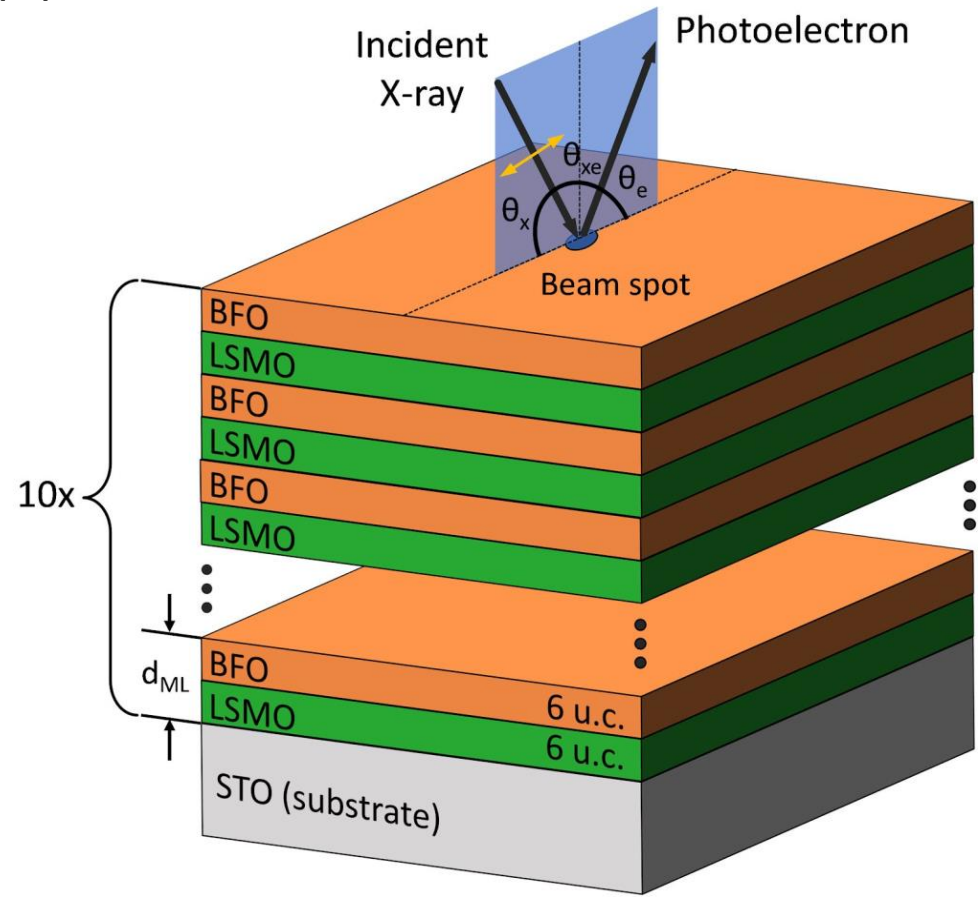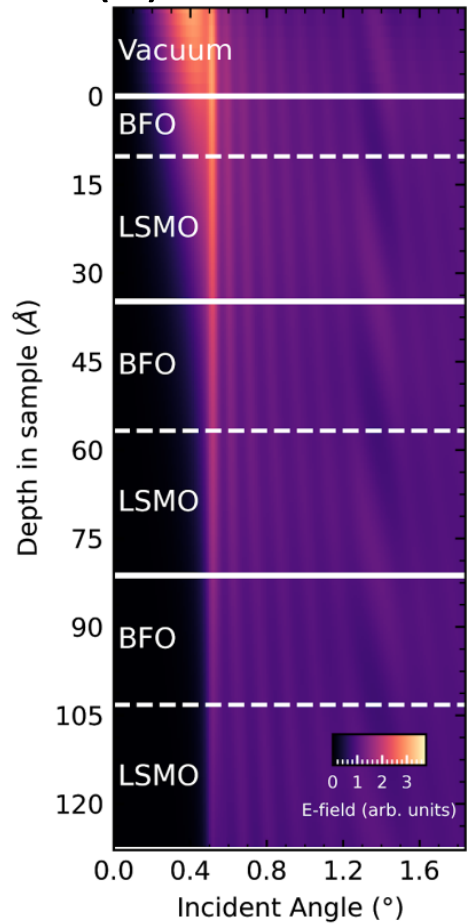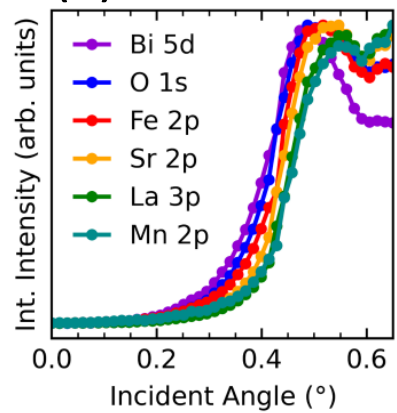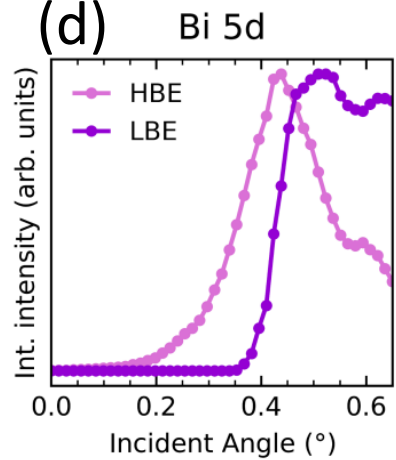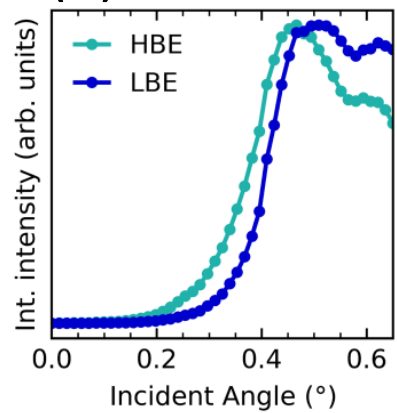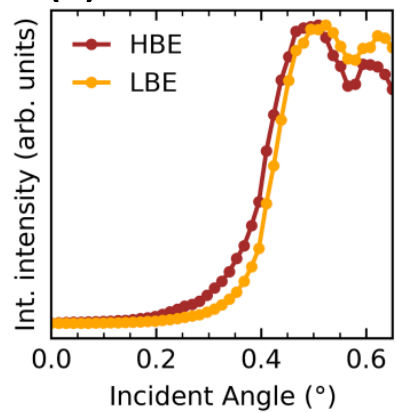

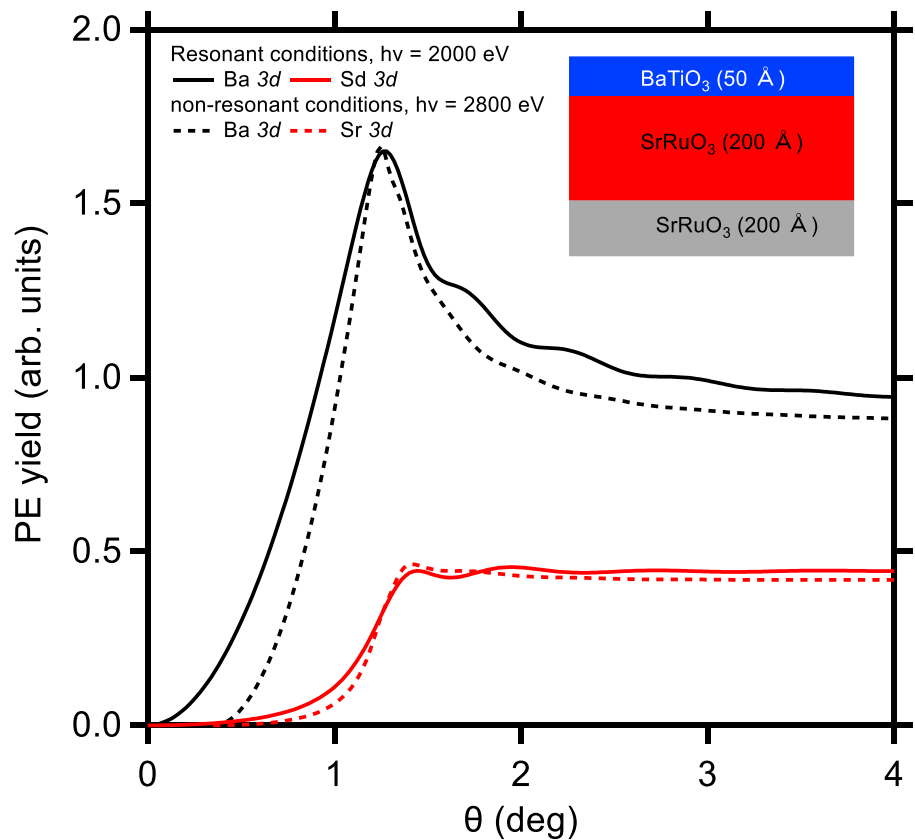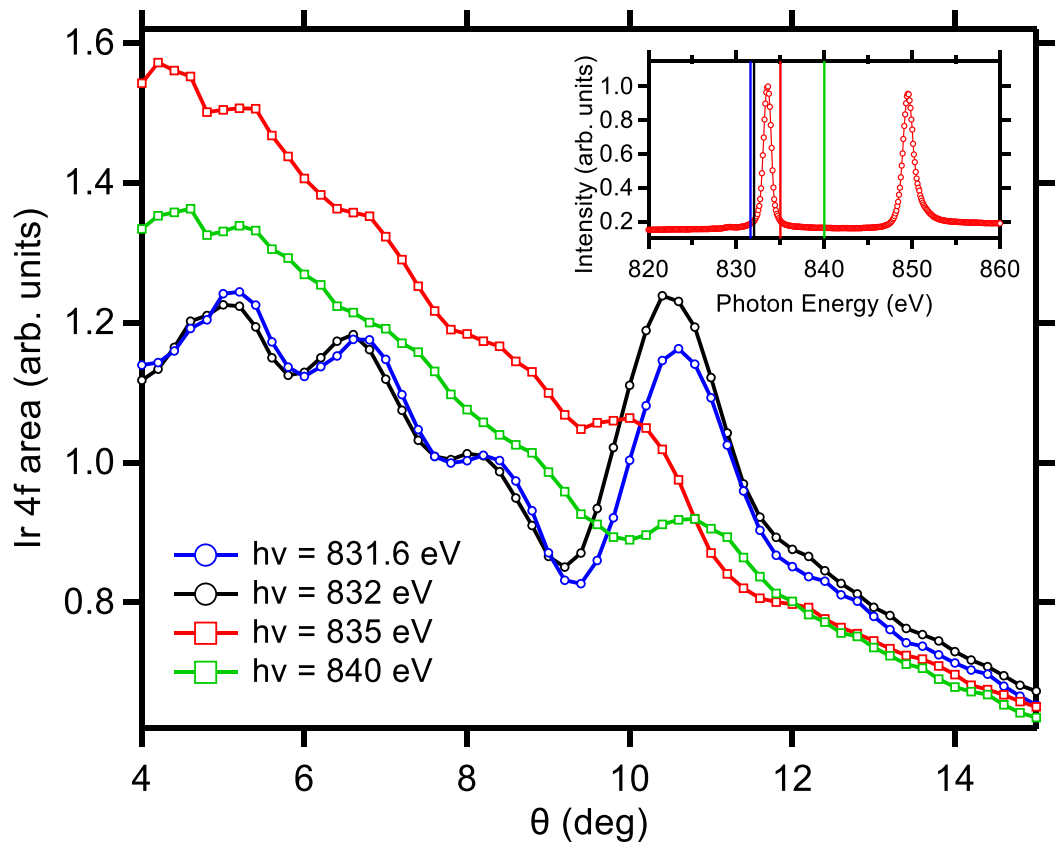